\documentclass[11pt,twoside]{article}
\usepackage{asp2010}
\resetcounters

\begin{document}

\title{A College-Level Inquiry-Based Laboratory Activity on Transiting Planets}
\author{Nicholas J. McConnell$^1$, Anne M. Medling$^2$, Linda E. Strubbe$^1$,
Pimol Moth$^3$, Ryan M. Montgomery$^2$, Lynne M. Raschke$^{4,5}$,
Lisa Hunter$^4$, and Barbara Goza$^6$\\
\affil{$^1$Department of Astronomy, University of California, 601 Campbell Hall, Berkeley, CA 94720; nmcc@astro.berkeley.edu and linda@astro.berkeley.edu}
\affil{$^2$Department of Astronomy \& Astrophysics, University of California, 1156 High St., Santa Cruz, CA 95064; amedling@ucolick.org}
\affil{$^3$Department of Astronomy, Hartnell Community College, 411 Central Ave., Salinas, CA 93901}
\affil{$^4$Institute for Scientist \& Engineer Educators and Center for Adaptive Optics, University of California, 1156 High St., Santa Cruz, CA 95064}
\affil{$^5$School of Sciences, College of St. Scholastica, 1200 Kenwood Ave., Duluth, MN 55811}
\affil{$^6$Department of Psychology, University of California, 1156 High St., Santa Cruz, CA 95064}}

\begin{abstract}
  We have designed an inquiry-based laboratory activity on transiting
  extrasolar planets for an introductory college-level astronomy
  class.  The activity was designed with the intent of simultaneously
  teaching science process skills and factual content about transits
  and light curves.  In the activity, groups of two to four students
  each formulate a specific science question and design and carry out
  an investigation using a table-top model of a star and orbiting
  planet.  Each group then presents their findings to other students
  in their class.  In a final presentation, the instructors integrate
  students' findings with a summary of how measured light curves
  indicate properties of planetary systems.  The activity debuted at
  Hartnell College in November 2009 and has also been adapted for a
  lecture-based astronomy course at U.C. Santa Cruz.  We present the
  results of student surveys before and after the astronomy course at
  Hartnell and discuss how well our activity promotes students'
  confidence and identity as scientists, relative to traditional lab
  activities.\end{abstract}

\vspace{-0.1in}
\section{Introduction}

We have designed an inquiry-based laboratory activity on transiting extrasolar planets for an introductory college-level astronomy class.
In our work, ``inquiry" means ``teaching science as science is done":  students learn about scientific concepts by figuring them out (as scientists do) instead of being given answers from a textbook or lecture.  Inquiry activities can model different parts of the scientific method: for example, students can design and carry out an entire investigation from designing the question to presenting results to others, or they can simply draw conclusions from a supplied dataset.  These sorts of activities enable students to learn content intertwined with research skills, like interpreting evidence, reasoning, thinking critically, conveying ideas, asking questions and figuring out how to answer them.  These skills are useful in later science classes, and also are important life skills.

Like real science, this technique can involve some winding pathways.  A student embarking on a self-designed investigation will likely hit dead ends or be confused by side issues, but with support he should eventually arrive at the desired conclusions.  Inquiry activities are not completely open-ended; teachers carefully monitor students' progress and help guide students toward desired conclusions through a process we call ``facilitation.''  Investigation with facilitator support allows the students to internalize the content they are studying; instead of words on a page, the results are activities they did and conclusions they figured out.  This process tends to be more engaging than following detailed step-by-step instructions or listening to a lecture, and studies have shown that the content learned lasts longer and is learned more thoroughly, as students have built up their own understanding instead of having it given to them \citep[e.g.][and references therein]{HSL, America}.

This activity was developed through the Professional Development Program (PDP) run by the Institute for Science and Engineer Educators (ISEE) at the University of California, Santa Cruz.  Each team of PDP participants designs an activity and facilitates its teaching for a specific venue.  These activities are ``backward-designed":  each team carefully takes into consideration the goals for the particular activity and the needs of the students first, and prioritizes elements of the activity based on these.  (For a more detailed discussion of the PDP, see Hunter et al., this volume; \citealt{PDP_description}.)  This activity was designed to fit in as one week of the semester-long introductory astronomy laboratory course (Astro 1L) at Hartnell Community College in Salinas, California.  Nicholas McConnell, Linda Strubbe, and Anne Medling were the primary designers and facilitators.  Pimol Moth is the instructor of Astro 1L.  Ryan Montgomery, Lynne Raschke, Lisa Hunter, and Barbara Goza provided additional support for the activity.

\section{The Venue and Activity}

Hartnell College is an accredited California Community College and Hispanic Serving
Institution located in Salinas, Monterey County, California, 120 miles south of
San Francisco. Of the College's 10,000+ students, 72\% are ethnic minority.  More than 40\% of the College's students are non-native English speakers, and 64\% are first-generation college students.  The majority of students enrolled in Astro 1L are from Hispanic backgrounds ($\sim$75-80\%), historically underrepresented in the sciences.  Astro 1L is predominantly taken by students majoring in non-science
subjects, who use the course to fulfill their physical sciences general education
requirement.

Astro 1L is a semester-long course that consists of 3-hour weekly lab sessions.
Fall 2009 was the first time that Hartnell College implemented inquiry in Astro 1L.  Two inquiry-based activities were included
in the Fall 2009 course: one on properties of lenses (Putnam et al., this volume),
and our activity, in which students investigate transiting
extrasolar planets.  Transiting planets are those that cause a periodic dimming in the light from their host star as they pass between the star and us on Earth.
The students learn to generate light curves (plots of the star's brightness over time)
and learn about the properties of the extrasolar planets by interpreting the trends in
the light curves.  Some important goals in Astro 1L are for students to gain an understanding of scientific processes, to view themselves as scientists, and to learn to interpret trends in data.   The hands-on knowledge gained in the transiting planets activity complements the information about extrasolar planets that is presented in the Introduction to Astronomy companion lecture class.

\section{Activity Goals}
\label{sec:goals}

Here we describe our goals for the activity and rationales for choosing them.  We mostly focused on goals that would have broad applicability to students' everyday lives: helping them to be curious, analytical, lifelong learners and helping them to communicate effectively with others.  In particular, we aimed to have students:
\begin{small}
\begin{enumerate}
\item devise their own questions about planetary systems and revise initial questions to form investigable ones;
\item construct a light curve (plot of brightness versus time) from their own measurements;
\item deduce relative properties of planetary systems from transit light curves (e.g., the planet's radius and orbital inclination);
\item present their work clearly and coherently; and
\item connect the content of the activity to current transit searches (e.g., the Kepler mission) and see that scientific discoveries are ongoing.
\end{enumerate}
\end{small}

We chose to emphasize the process of asking questions because of how broadly valuable this skill is in life.  We wanted to help students feel comfortable asking questions out loud in front of their peers.  In the classroom and beyond, asking questions helps students take charge of their own learning:  it pushes them to identify specific aspects they do not understand, thereby giving them an avenue toward finding out the answer.  Questioning can motivate students to pursue their curiosity, encouraging their ongoing learning about the world around them.  Furthermore, questions are the foundation of scientific inquiry and a crucial component of an authentic scientific experience.  We hoped that having students devise their own questions to investigate later would help give them ownership of the scientific content.

Our second and third goals above combined scientific content and processes.  We wanted students to get the scientific experience of taking their own data based on their own experimental set-up.  We then wanted the students to plot the results for several reasons:  to help them start to understand the connection between their model and their measurements, to understand why scientists make plots, and to see that their plots are essentially the same type that astronomers make to study actual transiting planet data.  The third goal
represents the heart of the scientific content:  students have to reason the same way astronomers do to understand the physical mechanism producing the different light curve shapes.  These two goals push students to make, interpret, and compare observations, and to connect their results to physical objects and processes in outer space.

Our fourth goal of developing students' presentation skills was chosen to help students deepen their understanding of the science, and for its broad value in life outside the classroom.
Students reinforce their comprehension by organizing their results mentally and visually on a poster, planning how to explain the results, vocalizing those explanations, and responding to questions about their explanations.  Learning to communicate ideas effectively can help students engage in and get what they need from their communities, including the one they create in the classroom.

Our final goal was to show students that scientific understanding is not static:  that our knowledge is constantly being tested and revised.  By connecting the activity to the current Kepler mission \citep{borucki}, we hoped to encourage students to follow real scientific discoveries in the news, and to feel empowered to understand them.  Sharing with family and friends would keep reinforcing their understanding about transiting planets.

\section{Activity Timeline}

{\bf Introduction} (25 minutes)
We began by introducing ourselves and verbally reminding students about what it means to participate in an inquiry-based lab, which can be difficult and frustrating but also rewarding.  We followed with a short slideshow presentation in which we described why astronomers study other planetary systems, defined a planetary transit and a light curve, showed examples of light curves, and stated that a light curve provides information about a planetary system.  We also mentioned the Kepler mission.

\begin{figure}[htbp]
\centering
\plotfiddle{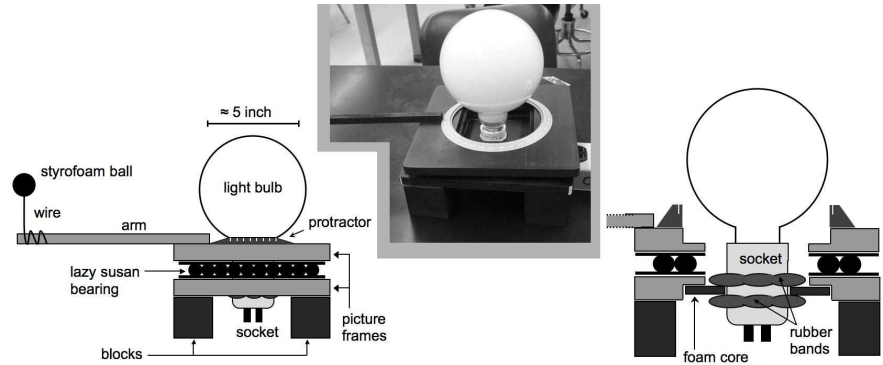}{1.5in}{0}{58}{58}{-173}{-173}      

\caption{\textit{Left:} Side view of the model orbital system built for the transiting planets activity.  All components except for the socket are held together with glue.  \textit{Middle:}  Photograph of one of the models used at Hartnell.  \textit{Right:}  Cut-away view of the model orbital system.  The light bulb socket is fed through a hole cut into a foam core mount.  Before the light bulb is attached, rubber bands are added on either side of the hole to hold the socket in place.}
\label{fig:model}
\end{figure}

\medskip
\noindent {\bf Questioning} (30 minutes)
We began by introducing table-top model planetary systems (built by NM, LS, and AM; see Figure~\ref{fig:model} and Appendix~\ref{app}) and asked the students to spend ten minutes playing with the models and additional materials.  This was to familiarize the students with the available materials, and to establish a safe atmosphere for free experimentation and brainstorming.  When a few minutes remained, we interjected with suggestions if students seemed to be overlooking particular variables (e.g. orbital inclination).  Additionally, we provided a handout containing words like ``transit,'' ``brighter,'' and ``twice as big'' for inspiration if the students felt stuck with devising their own questions.

The next steps followed:
students individually brainstormed on paper their first impressions and ideas about transiting planets;
facilitators described the idea of refining impressions into specific investigable questions, and went through one example;
students individually refined their first impressions;
small groups continued refining, and selected their favorite questions;
groups shared their favorite questions with the class; and
instructors classified the questions as ``Investigable Today'' and ``Not Equipped'' (respectively, questions which could and could not reasonably be addressed with the model planetary systems).

With these steps, we aimed to build up students' abilities to devise specific, investigable questions, and to help students feel ownership of the questions they would ultimately investigate.  It was crucial to demonstrate to students that all proposed questions were valuable, regardless of whether they were ultimately eligible for investigation.  As students shared their questions with the class, we had an opportunity to assess how well we were achieving Goal \#1.

\medskip
\noindent {\bf Investigations} (60 minutes)
We asked each student group (2-4 students) to select a question from the ``Investigable Today'' category.  Next, we demonstrated how to use a digital light meter to measure brightness during a model transit, and demonstrated plotting those data as a light curve.  We then instructed students to use the model planetary system, light meter, and additional materials to investigate the answer to their question.

By eavesdropping on and occasionally questioning students' investigations, we were able to assess their understanding of the relationship between their experimental set-ups and the resultant light curves they plotted (Goals \#2 and \#3).  In particular, we wanted students to use the light curves as a tool to gain insight about transit phenomena, instead of regarding the light curves as the final product of their investigation.  We looked for students to recognize the connection between unresolved, quantitative outputs of the light meter and a resolved image of the star's surface partially blocked by a planet.  One common facilitation technique was to ask a student to replace the light meter with her eye and describe what she saw.

\medskip
\noindent {\bf Presentations} (45 minutes)
Still in their groups, students prepared and gave oral poster presentations on the results of their investigations (Goal \#4).  Each group had three minutes to present their poster and findings, and three minutes for questions from the audience.  We required each audience group to ask at least one or two questions per presentation (Goal \#1), and offered a list of suggested questions if they needed inspiration.  The presentations gave us another opportunity to assess students' understanding of how light curves can be used to deduce properties of planetary systems (Goals \#2 and \#3).

\medskip
\noindent {\bf Synthesis} (10 minutes)
We concluded with a slideshow presentation to recap the activity.  We described the ``Thinking Skills'' students had worked on: asking questions (Goal \#1), explaining their ideas to each other (Goal \#4), and designing and carrying out an experiment.  We also described how different physical properties of planetary systems lead to different observed light curves, in order to reinforce students' understanding of the data and the physics (Goals \#2 and \#3).  We finished by reminding them about current transit searches like the Kepler mission, and pointed them to two recent popular-level science articles on transits (Goal \#5).

\section{Instructor Reflection and Student Feedback}

The activity at Hartnell was overall very successful.  Many students met many of our goals, and 90\% of students wrote positive responses on their anonymous post-activity surveys.

The students built their own investigable questions (Goal \#1) quite successfully.  We did not ask to see their impressions or first attempts at questions, so unfortunately we could not watch the questions develop, but ultimately each group was able to share at least one relevant, coherent question with the class.  Some examples of students' investigable questions included the following and varieties thereof:
\begin{small}
\begin{itemize}
\item ``How can the size of the planet be determined?''
\vspace{-0.05in}
\item ``How does the brightness of the star affect how well we can detect the planet?''
\vspace{-0.05in}
\item ``In what ways does the inclination of a planet affect the transit?'' and
\vspace{-0.05in}
\item ``Can you tell if a planet has a ring?''
\end{itemize}
\end{small}
Students also asked questions for which we were not equipped to support an investigation, like:
\begin{small}
\begin{itemize}
\item ``Does the size of a planet affect its orbital period around the star?'' and
\vspace{-0.05in}
\item ``Does the size of our telescope affect our ability to view the planet?''
\end{itemize}
\end{small}

A majority of groups selected questions related to planet size or orbital inclination, but a sizeable fraction selected other questions related to planetary rings, atmospheres, or reflected light from planets.

The student groups generally worked well together on their investigations.  Often, each student had a role in the measurement-taking process (e.g., one moved the model planet, one used the light meter, one recorded the brightness).  Most students successfully plotted at least one light curve based on their group's measurements (Goal \#2).  Brightness measurements were easy for most, but many had difficulty in using the equipment to measure the phase angle, or in keeping a steady orbital rate to measure even time increments.  This could be improved for future labs.

Some students were able to deduce relative planetary properties from light curves (Goal \#3).  The most successful tended to be those who studied planetary size.  They could generally explain that a larger planet blocks more light and therefore produces a larger dip in the light curve; only rarely, though, did a student turn this around and articulate that astronomers can measure planet sizes from light curves.  Groups who studied orbital inclination usually realized that some planetary systems will not show transits to a given observer, and often realized that this fraction of systems is large.  The groups studying rings, atmospheres, and reflected light did not tend to demonstrate strongly meaningful results.  (The model planetary systems' limited ability to accurately depict these phenomena likely contributed to the students' difficulties in these studies.)

Students' presentations to their classmates showed room for improvement.  Most posters did show a few light curves and diagrams of the experimental set-up or face of the star during transit.  Yet most students described the various steps and wrong turns that they took, with little aim at providing final results and explanations.  Due to time constraints, we did not offer much support for presentation preparation:  only a quick skeleton of guidelines and 10-15 minutes to prepare.  We successfully elicited questions from each group to ask the presenters, although the audience was often reluctant and relied upon the list of suggested questions we provided.  When students asked their own questions, they were usually related to the asker's own experiment (e.g., ``Did you try putting in different sized planets?''), which indicates the asker's ownership of their subtopic.

As mentioned above, 90\% of students wrote something positive on their feedback forms.  A large majority used the words ``interesting'' and/or ``fun'' to describe the activity.  Students said that they liked working with the model and the light meter, they liked asking their own questions and designing their own experiments, and they liked working in groups.  The most common negative comment was that students felt that they did not have enough time during the activity.  Students also said that the lab was difficult or frustrating, some wished for more help and guidance from us, and some felt uncomfortable by our ``hovering'' as they worked.

A few sample student comments are:
\begin{small}
\begin{itemize}
\item ``I really liked that we had to think about how were going to answer the question that we chose; even though it is frustrating it feels good to think like that.'' 
\item ``It was a little difficult to concentrate for me personally because we kept getting checked on and asked what our solution was when we ourselves didn't know at the time.'' 
\item ``It made us think like real astronomers, build our own question; hypothesis and make our own data.'' 
\item ``Because it was our own experiment, I felt like I was a scientist.'' 
\end{itemize}
\end{small}

\section{Comparison of Pre- and Post-Semester Surveys Before and After Inquiry}

In order to assess the effectiveness of introducing inquiry into the laboratory, we conducted a survey of students enrolled in Astro 1L during the Spring 2009 semester (without inquiry) and during the Fall 2009 semester (with inquiry).  Adapted from earlier research \citep{chemers}, the survey assessed students' levels of \emph{self-efficacy for science}, \emph{identity as a scientist}, and \emph{commitment to a science career}.  In both semesters, students anonymously completed the survey during the first week of classes (pre-semester) and again at the end of the semester (post-semester).   We expected students involved in the inquiry-based activities to report greater gains in confidence in their science skills and their interest in pursuing a career in science.

In the \emph{self-efficacy for science} construct, students responded to 13 declarative statements such as ``I am confident I can generate a research question to answer" on a scale from 1 (``not confident at all") to 5 (``absolutely confident").  For \emph{identity as a scientist} (5 items) and \emph{commitment to a science career} (7 items), students responded on a scale from 1 (``strongly disagree") to 5 (``strongly agree").  Initial analyses determined that the constructs had good psychometric properties, with a single factor and high internal consistency.  The descriptive statistics are found in Table 1.  Independent $t$-tests conducted to compare means show that there was a statistically significant increase in all of the post-semester constructs ($t > 2.60$) except for Fall \emph{commitment to a science career}.  This implies that students in general were more confident in their science abilities, identified more as scientists, and were more committed to a science career at the end of both semesters.   There is, however, not a statistically significant difference in the means between the two semesters.  Given that during the period of this assessment, only two inquiry activities were introduced out of 18 total labs, we cannot reliably deduce the level of effectiveness of adding inquiry to the laboratory curriculum.  Furthermore, because these were anonymous surveys, we cannot track changes for individual students. In the future, we will be better able to assess this when more inquiry activities are introduced, the data are analyzed over several semesters, and we track individual students' responses across the semester. We expect to find a subset of students for whom the inquiry laboratories inspire greater interest in science and greater confidence about participating in science.
\begin{table}[!ht]
\label{survey}
\caption{Results of surveys to analyze the effect on students of adding inquiry activities to the course.  The `**' indicates a statistically significant difference between pre- and post-semester means.  `SD' refers to the standard deviation of student responses, where `n' is the number of student responses included in the statistics.}
\smallskip
\begin{center}
{\small
\begin{tabular}{cccc|cccc}
\tableline
& \multicolumn{3}{c}{\textbf{Pre-semester}} & \multicolumn{3}{c}{\textbf{Post-semester}} & \\
\textbf{Spring 2009 (pre-inquiry)} & Mean & SD & n & Mean & SD & n & Indep \textit{t}  \\
\tableline
\noalign{\smallskip}
Self-Efficacy for Science & 3.01 & 0.73 & 62 & 3.51 & 0.72 & 47 & -3.55**  \\
Identity as a Scientist & 2.07 & 0.82 & 69 & 2.98 & 0.92 & 63 & -5.98** \\
Commitment to a Science Career & 2.11 & 0.99 & 69 & 2.60 & 1.17 & 63 & -2.66** \\
\tableline
\textbf{Fall 2009 (post-inquiry)}  & Mean & SD & n & Mean & SD & n & Indep \textit{t} \\
\tableline
\noalign{\smallskip}
Self-Efficacy for Science & 2.73 & 0.87 & 65 & 3.40 & 0.85 & 56 & -4.26** \\
Identity as a Scientist & 2.12 & 0.92 & 68 & 2.72 & 0.94 & 61 & -3.68**\\
Commitment to a Science Career & 2.14 & 1.05 & 68 & 2.36 & 1.10 & 59 & -1.12\\
\tableline
\end{tabular}
}
\end{center}
\end{table}

\section{Suggestions for Future Implementation}

We designed our activity to meet our goals within the time
constraints. Here, we suggest modifications to the activity for
different time constraints or to achieve different goals.

Teachers may choose to focus the activity more on understanding the
differences between models and the physical systems they represent.
Because it is impossible on a tabletop to accurately represent the
astronomical distances in a solar system, students may be misled by
the scale of the models and, in particular, not realize that transit
searches are necessary because we cannot resolve the planet separately
from the star.  To address these concerns, students should explicitly
consider the fact that they are using a model, and could discuss other possible models.  These may include scaling for distances: e.g., a light bulb in the classroom (Sun), a marble in the parking lot (Jupiter), and a second light bulb a few thousand miles away (the nearest star).  To address relative brightnesses, students can discuss a model in which a bright lightbulb (Sun) is next to a peppercorn (exoplanet).  To further emphasize that astronomers
cannot observationally resolve planets, one might add wax paper to one end of the model ``telescopes" (toilet paper
rolls).  Students then cannot see the planet but should notice that
the light gets dim periodically, and can then begin a discussion of light
curves.  A computer model could also be discussed.

Teachers may choose to focus further on the process of questioning.
The facilitators could spend more time discussing what makes a
question ``investigable", and the students could even categorize
proposed investigation questions rather than the facilitators.  To
help students generalize the questioning process to other aspects of their
lives, students could read a short article from a Voter Information
Packet, then discuss questions that they had and how they might find
the answers.

Building presentation skills is another direction to focus.  One of
the simplest ways is to give students more time to prepare their
presentations.  Another idea is to discuss explicitly what constitutes
a good explanation, giving students a framework to rely on or
a template to follow.  Facilitators could also lead a class discussion
about what goes into an effective presentation.  It would be helpful
to give students an opportunity to practice their presentations before
getting up in front of the class.  If presentation
skills were a focus for an entire semester-long course, students could
likely improve significantly by giving presentations every week.  Whether
students give presentations frequently or only once, it is worth giving
each student or group detailed feedback when possible.

A final suggestion is to focus more on awareness of current
scientific research.  Reading a
popular science article, discussing it with classmates or as a class,
and presenting a summary of it are all good ways to get students more confident and familiar with talking about current scientific research.

\section{Implementation in a Lecture Course}
Here we describe specific modifications to the transiting planet inquiry for a different teaching environment (performed in Spring 2010):  U.C. Santa Cruz's introductory astronomy course for non-majors (Astronomy 2), a large (250 student), lecture-based survey course.
In order to give students practice with scientific process skills, instructor Ryan Montgomery adapted the activity to be completed during one of the 70-minute (required) discussion sections.  Two Teaching Assistants (TAs) assisted the students in each discussion section of $\sim$40 students.

The aim was to have students work outside of discussion section to complete segments of the inquiry activity that required little or no facilitation, maximizing the utility of the TAs during section.  A website gave a brief introduction to the scientific content (the transit detection method) and showed a series of demonstration videos to familiarize the students with the model planetary systems.  The website then asked student groups to complete a pre-lab assignment of generating questions based on the video demonstrations.
The pre-lab group activity was to be completed and turned in by each group to their TA at least 24 hours prior to their discussion section, providing time to sort and electronically post the questions. Before arriving at section, groups were to decide on a question that they wanted to investigate.
The discussion was then used solely for investigation, with a brief ($\sim$10 minute) sharing/synthesis segment at the end of the period.  Formal student presentations were cut from this implementation; the goal of having students present their work was met later in the course, when they gave formal presentations as part of a different activity.

Overall this version of the inquiry activity was well received, and met the course content and process goals.  Students were asked to rate the amount of content they learned, opportunities to practice the processes of science, their enjoyment, and the activity overall.  On a scale from 1 to 5, students consistently rated the activity at 3.6, between ``Fair" (3) and ``Good" (4).  By later comparing students' ratings of the inquiry activity with their post-course ratings of their TAs, we conclude that TA support is likely responsible for some of the overall effectiveness of the transiting planet inquiry.
We believe that the modified activity successfully retained the self-direction, ownership, and engagement that make inquiry activities valuable.  We strongly encourage other lab and lecture-based courses to consider using an implementation of this transiting planet inquiry activity. 

\acknowledgements The authors acknowledge the National Science Foundation Science and
Technology Center funding of the Center for Adaptive Optics, managed by the
University of California, Santa Cruz, under cooperative agreement No.
AST-9876783.  This work was funded in part by the National Science Foundation,
through the Course, Curriculum and Laboratory Improvement program
(DUE \#0816754), and supported by the U.C.~Santa Cruz Institute for Scientist \& Engineer Educators.  We thank the U.C.~Santa Cruz Physics Department for allowing us to borrow light meters for the activity.

\appendix
\section{Building a Turn-Table Model of a Planetary System}
\label{app}

Below we list the required materials for one model, with part numbers where applicable.  Figure~\ref{fig:model} depicts how the parts are assembled to make a model.  The costs enumerated below add to about \$35.  Some supplies (foam core, styrofoam balls, glue, spray paint) can be used to assemble multiple models; with these supplies already purchased, a second model would cost about \$22.  For further instructions to assemble the model, we encourage readers to contact NM, AM, or LS.\\

\begin{small}
\begin{itemize}
\vspace{-0.05in}
\item 2 wooden picture frames (Michael's Crafts, SKU 400100118634, \$1.00 each)
\vspace{-0.05in}
\item lazy susan bearing (The Home Depot, SKU 039003095485, \$4.49)
\vspace{-0.05in}
\item globe light bulb (The Home Depot, SKU 043168908320, \$4.27)
\vspace{-0.05in}
\item light bulb socket with male 2-prong end (The Home Depot, SKU 078477077375, \$2.09)
\vspace{-0.15in}
\item 3 cubic wooden blocks (2 inches) (Michael's Crafts, SKU 754246103147, \$1.29 each)
\vspace{-0.05in}
\item plastic circular protractor (Michael's Crafts, SKU 79252360026, \$2.99)
\vspace{-0.05in}
\item long rectangular wooden craft stick (Michael's Crafts, \$1.59)
\vspace{-0.05in}
\item foam core board  (Michael's Crafts, SKU 79946129960, \$2.99)
\vspace{-0.05in}
\item styrofoam or clay ball  (Michael's Crafts, various sizes, $\sim$\$4 for 6-12 styrofoam balls)
\vspace{-0.05in}
\item super glue or wood glue  (super glue: Michael's Crafts, SKU 70158009255, \$3.29)
\vspace{-0.05in}
\item black spray paint, flat finish (The Home Depot, SKU 020066187811, \$3.44)
\vspace{-0.05in}
\item wire
\vspace{-0.05in}
\item thick rubber bands (2 to 4)
\end{itemize}
\end{small}

\noindent Digital light meter prices range from \$30 to \$300 each (an example in the \$100 range is the DLM 1337 from General Tools).  Although we used one light meter per group at Hartnell, we believe that students could investigate transits successfully with one light meter per two groups.  Medium-sized styrofoam balls block approximately $10\%$ of the light bulb's surface; we recommend using light meters that are sensitive enough to reliably detect $10\%$ contrast from a few meters away.
\bibliography{mcconnelletal_refs}

\end{document}